\documentclass[final,3p,times,twocolumn]{elsarticle}

\usepackage{graphicx}
\usepackage{xcolor}
\usepackage[caption=false,justification=justified]{subfig}  

\usepackage{amssymb}
\PassOptionsToPackage{hyphens}{url}\usepackage{hyperref}
\usepackage{listings}
\usepackage{amsmath}
\usepackage[autoload=false]{jlcode}
\usepackage{xcolor}
\usepackage[normalem]{ulem}
\usepackage{hyperref}
\hypersetup{
    colorlinks=true,
    linkcolor=blue,
    citecolor=blue,
    urlcolor=blue,
    filecolor=blue,
    pdfborder={0 0 0},
    breaklinks=true
}

\definecolor{codegreen}{rgb}{0,0.6,0}
\definecolor{codegray}{rgb}{0.5,0.5,0.5}
\definecolor{codepurple}{rgb}{0.58,0,0.82}
\definecolor{backcolour}{rgb}{0.95,0.95,0.92}
\definecolor{darkGreen}{RGB}{0,110,0}

\lstdefinestyle{mystyle}{
  backgroundcolor=\color{backcolour}, commentstyle=\color{codegreen},
  keywordstyle=\color{magenta},
  numberstyle=\tiny\color{codegray},
  stringstyle=\color{codepurple},
  basicstyle=\ttfamily\scriptsize,
  breakatwhitespace=false,         
  breaklines=true,                 
  captionpos=b,                    
  keepspaces=true,                 
  numbers=none,                    
  numbersep=5pt,                  
  showspaces=false,                
  showstringspaces=false,
  showtabs=false,                  
  tabsize=2,
  frame=single,
  framexleftmargin=-5mm,
  framexrightmargin=+0mm,
  framextopmargin=1mm,
  framexbottommargin=-3mm,
  xleftmargin=-.4cm,
  xrightmargin=+0.08cm,
  escapeinside={(*@}{@*)},
}

\usepackage{float}

\newcounter{bla}

\makeatletter
\def\ps@pprintTitle{%
  \let\@oddhead\@empty
  \let\@evenhead\@empty
  \def\@oddfoot{\reset@font\hfil\thepage\hfil}%
  \let\@evenfoot\@oddfoot
}
\makeatother

\newcommand{\beforeit}[0]{{\fontfamily{cmss}\selectfont
BeforeIT}}

\newcommand{\modelname}[0]{{\fontfamily{cmss}\selectfont{base model}}}

\newcommand{\dfont}[1]{{\fontfamily{lmss}\selectfont
#1}}

\newcommand{\code}[1]{\texttt{#1}}

\usepackage{flushend}

\usepackage{lipsum}

\begin{document}

\begin{frontmatter}





\title{\texorpdfstring{\vspace{-2cm}}{} \dfont{BeforeIT.jl}: High-Performance Agent-Based Macroeconomics Made Easy}


\author[a]{Adriano Meligrana}
\author[b,c,d]{Mitja Devetak}
\author[e]{Steven Hoekstra}
\author[f]{Oliver Reiter}
\author[g]{Enrico Minnella}
\author[h,i]{Sebastian Poledna}
\author[g,j]{Marco Bardoscia\texorpdfstring{$^{\dagger}$}{}}
\author[k]{Aldo Glielmo\texorpdfstring{$^{\ddagger}$}{}}

\address[a]{Sapienza University of Rome, Rome, Italy}
\address[b]{Paris 1 Panthéon-Sorbonne University, Paris, France}
\address[c]{Complexity Science Hub, Vienna, Austria}
\address[d]{Supply Chain Intelligence Institute Austria, Vienna, Austria}
\address[e]{University of Amsterdam, Amsterdam, The Netherlands}
\address[f]{Vienna Institute for International Economic Studies (wiiw), Vienna, Austria}
\address[g]{Bank of England, London, United Kingdom$^{\ast}$}
\address[h]{University for Continuing Education Krems, Krems, Austria}
\address[i]{Austrian Institute of Economic Research (WIFO), Vienna, Austria}
\address[j]{University College London, Department of Computer Science, London, United Kingdom}
\address[k]{
Banca d’Italia, Directorate General for Information Technology, Applied Research Team, Rome, Italy$^{\ast}$}
\nonumnote{$^{\dagger}$marco.bardoscia@bankofengland.co.uk}
\nonumnote{$^{\ddagger}$aldo.glielmo@bancaditalia.it}

\nonumnote{$^{\ast}$The views and opinions expressed in this paper are those of the authors and do not necessarily reflect the official policy or position of Banca d’Italia or those of the Bank of England or its committees.}

\begin{abstract}

\beforeit{} is an open-source software for building and simulating state-of-the-art macroeconomic agent-based models (macro ABMs) based on the recently introduced macro ABM developed in \cite{poledna2023economic} and here referred to as the \emph{\modelname{}}.
Written in Julia, it combines extraordinary computational efficiency with user-friendliness and extensibility.
We present the main structure of the software, demonstrate its ease of use with illustrative examples, and benchmark its performance.
Our benchmarks demonstrate that the \modelname{} built with \beforeit{} is orders of magnitude faster than its original Matlab implementation, while achieving performance comparable to a highly optimised HPC implementation.
Furthermore, we show how \beforeit{} supports an easy construction of model variants such as the `CANVAS' model of \cite{hommes2025canvas}, how these variants can be evaluated at scale via an automated multicountry forecasting pipeline, and how the entire simulation can be automatically differentiated.
\beforeit{} is designed to facilitate reproducibility, extensibility, and experimentation.
As the first open-source, industry-grade software to build macro ABMs, \beforeit{} can significantly foster collaboration and innovation in the field of agent-based macroeconomic modelling. 
The package, along with its documentation, is freely available at \url{https://github.com/bancaditalia/BeforeIT.jl} under the Apache 2.0 license.

\end{abstract}

\end{frontmatter}

\newpage

\section{Introduction}

%
Macroeconomic agent-based models (macro ABMs) are computational models in which the micro-level behaviour of heterogeneous interacting agents, such as households and firms, is simulated to produce a macro-level understanding of aggregate economic outcomes \cite{hamill2015agent,dawid2018agent,axtell2025agent}.
%
This modelling approach is in contrast with traditional macroeconomic frameworks based on dynamic stochastic general equilibrium (DSGE) models, which solve for equilibrium conditions under well-defined optimisation problems \cite{christiano2010dsge,del2013dsge,christiano2018dsge} and typically, but not always \cite{gali2007understanding,kaplan2018monetary}, use representative agents.
In comparison to DSGE models, macro ABMs use computational simulations of heterogeneous agents interacting, often requiring greater computational resources and more extensive calibration \cite{pangallo2024data}.
On the other hand, ABMs can naturally treat heterogeneity and nonlinear interactions among agents, as well as nonequilibrium dynamics.

Since their emergence in the early 2000s, macro ABMs have become an important tool for analysing economic dynamics, and particularly for reproducing stylised facts and capturing emergent phenomena such as asset bubbles, market crashes, economic divergence and convergence, and technological innovation \cite{turrell2016agent,dawid2018agent,dosi2019more,dawid2024implications}. 
ABMs can also provide a strong foundation for policy analysis in areas such as systemic risk, housing market dynamics or inflation, and they have been increasingly employed by policy institutions \cite{turrell2016agent,catapano2021macroprudential,carro2023heterogeneous,bardoscia2025impact,axtell2025agent,minnella2026devil}.

\begin{figure}[t]
\centering
\includegraphics[width=0.99\linewidth]{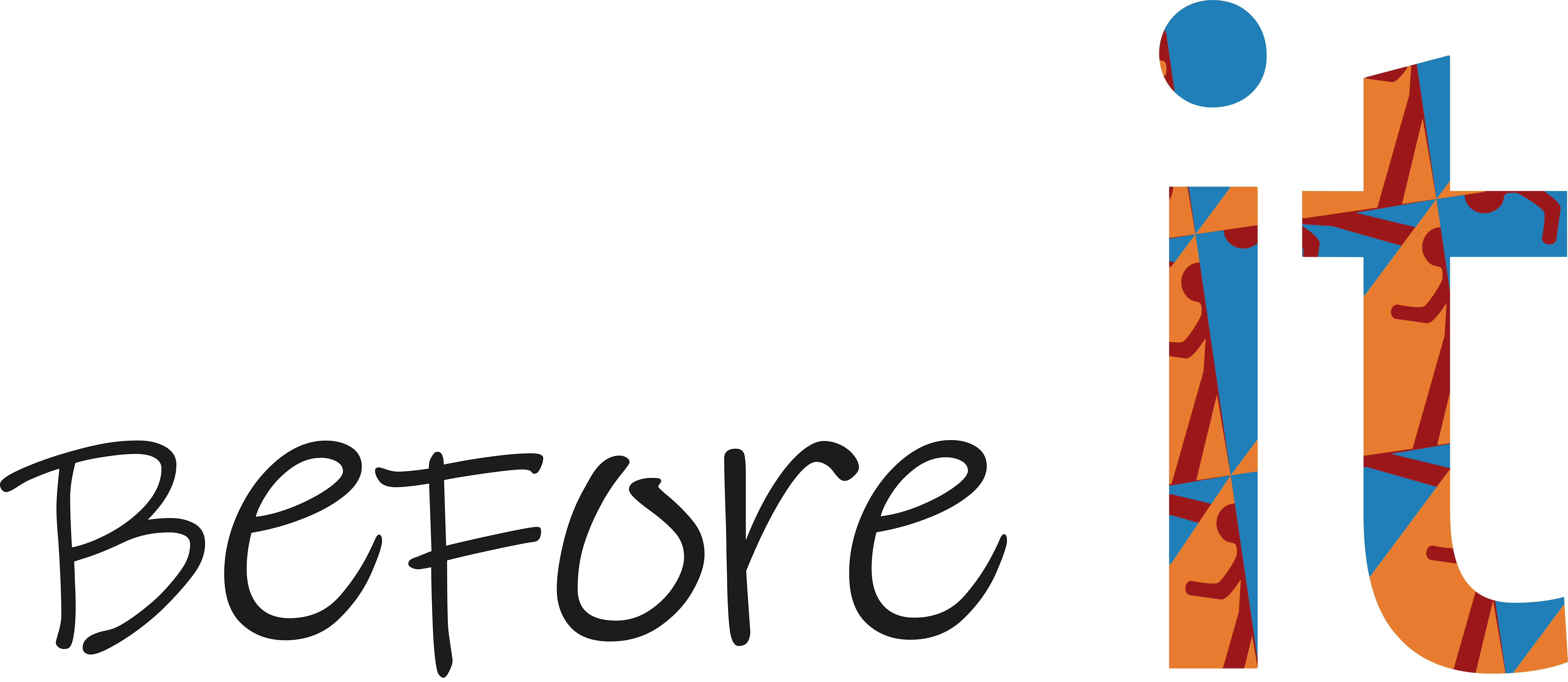}
\caption{{\bf Logo of the package.}
\beforeit{} stands for \textbf{B}ehavioural \textbf{e}conomic \textbf{fore}casting by the \textbf{IT} research unit of the Bank of \textbf{It}aly. The agents in the `it' monogram exemplify the agent-based nature of the simulation method.
The name of the package also refers to the forecasting possibilities offered by the package, as in predicting a phenomenon \emph{before it happens}.
}
\label{fig:logo}
\end{figure}

%
%
The macro ABM developed in \cite{poledna2023economic} represents an important step forward for the field as it demonstrated that ABMs, alongside their capacity to reproduce emergent phenomena and stylized facts, can also achieve short-term forecasting performances competitive with vector autoregressive (VAR) and DSGE models.
The model’s flexibility and detailed representation of economic sectors has enabled it to address a variety of issues, ranging from the COVID-19 pandemic \citep{poledna2020recovery} and the 2021–2023 inflation surge \citep{grazzini2023understanding} to financial crises \citep{hommes2023analyzing}, migration \citep{poledna2024economic}, natural disasters \citep{bachner2024revealing,hochrainer2024risk}, and labour productivity losses due to heat stress \citep{kimmich2024economic}. 
The model was extended to work in distributed memory parallel environments in \cite{gill2021high}, and was taken as the basis for the multi-country macro ABM developed in \cite{wiese2024forecasting}.
Interest in this model has grown among several institutions, including the Bank of Canada, which incorporated it into monetary policy analysis \citep{hommes2025canvas,coletti2023blueprint,gosselin2023making}.

Over the years, this broad adoption has given rise to multiple implementations of such a model. 
The original implementation, available in open source, is written in Matlab \cite{poledna2023economic,iiasa-model-github},
and it has served as the foundation for other variants. 
For instance, the Bank of Canada created CANVAS --— a production-oriented modification of the reference Matlab code --— which is currently closed-source. 
While Matlab is a closed-source language, the original open-source Matlab code can also be run using the open-source Octave language, although this typically implies a significant slowdown in execution time.
A Distributed Memory Parallel (DMP-HPC) implementation in C++ using MPI for supercomputers, which is likewise closed-source, has been introduced in \cite{gill2021high}. 
Meanwhile, the University of Oxford \cite{wiese2024forecasting} and Deloitte \cite{simudyne} have each developed their own closed-source extensions in Python and Java, respectively, underscoring the model’s adaptability across a range of environments.

Despite this array of implementations, until now, there was no user- and developer-friendly package that could be freely accessed, installed, and extended for diverse applications. 
To address this gap, we developed and released \beforeit{}, an open-source, thoroughly documented software package that is both user- and developer-friendly, providing computational efficiency and a structure suitable for extensions. 
\beforeit{} takes the model developed in \cite{poledna2023economic} as its core reference, but it also offers a general platform that allows researchers to construct, modify, and extend macro ABMs tailored to their specific applications, thus fostering methodological experimentation and custom analysis. 
Since the model in \cite{poledna2023economic} forms the foundation of \beforeit{}, we refer to it as the \modelname{} in the rest of this work.

To ensure quality, reliability, and long-term usability, we made \beforeit{} adhere to modern open-source package development standards and best practices.
\beforeit{} is publicly available on the GitHub page of the Bank of Italy\footnote{\href{https://github.com/bancaditalia/BeforeIT.jl}{https://github.com/bancaditalia/BeforeIT.jl}}, expanding the organization’s library of open-source ABM software packages, which includes other models \cite{brusatin2024simulating} and a calibration engine \cite{benedetti2022black,glielmo2023reinforcement}.
Its name and logo, illustrated in Fig.~\ref{fig:logo}, highlight its focus on behavioural agent-based economic forecasting.

Developed in the Julia language, \beforeit{} achieves computational efficiency comparable to C-compiled code while preserving readability and maintainability, characteristics typically associated with scripting languages such as Python or Matlab. 
The overall design of \beforeit{} is, in part, inspired by Agents.jl~\cite{datseris2024agents}, the leading general-purpose agent-based modelling framework in Julia, from which we borrow the idea of a transparent, function-based stepping interface.
Julia's balance between ease of use and computational performance has contributed to Julia's growing popularity in economics for numerical computations \cite{Dynarejl,kockerols2023macromodelling}, with notable adoptions by institutions such as the Federal Reserve \cite{del2015frbny,del2017forecasting} and the Bank of Canada \cite{StateSpaceEconjl,araujo2024open}. 

By providing a platform to build, run and extend a model of proven impact, \beforeit{} fills an important gap in the open-source software community with the aim of fostering innovation, collaboration, and the wider adoption of macro ABMs.

The rest of this work is structured as follows.
In Sec.~\ref{sec:model-summary} we briefly review the \modelname{} as this forms the foundation of \beforeit{}, in Sec.~\ref{sec:software-structure} we describe the structure of the software and its main features, in Sec.~\ref{sec:usage-illustration} we provide practical examples of software usage, starting from an essential script and moving to more advanced usage such as for scenario analysis, model calibration and automatic differentiation, in Sec.~\ref{sec:model_improvements} we show how \beforeit{} enables the systematic definition and large-scale evaluation of model variants, finally in Sec.~\ref{sec:conclusions} we conclude.

\section{An essential model summary}
\label{sec:model-summary}
The \modelname, the macro ABM developed in \cite{poledna2023economic}, simulates macroeconomic dynamics in a small open economy by modelling the interactions of millions of heterogeneous agents. 
%
%
The model is calibrated using data from national accounts, sectoral data, input-output tables, and business demographics, ensuring that the model reproduces exactly the state of the economy in that quarter in terms of aggregate GDP, GDP components, and industry sizes.

The model adheres to the European System of Accounts (ESA) framework \cite{ESA2010} and organises the economy into the following six classes of macroeconomic agents:
\begin{itemize}
    \setlength\itemsep{-0.15em}
    \item \textbf{Households} make consumption, savings, and labour supply decisions.
    \item \textbf{Non-financial corporations} produce goods using labour, capital, and intermediate inputs, with firm-level heterogeneity captured in terms of size, production processes, and market conditions. 
    \item \textbf{Financial institutions} mediate credit markets.
    \item \textbf{The central bank} implements monetary policies.
    \item \textbf{The government} is responsible for taxation, public goods provision, and income redistribution
    \item \textbf{The rest of the world} accounts for external trade and international capital flows.
\end{itemize}

While in the \modelname{} the financial sector consists of a single representative bank, \beforeit{} generalises it to an arbitrary number of banks, set by the parameter \code{n\_banks}.
Each firm and household is randomly assigned to one bank, and each bank keeps its own balance sheet and extends credit subject to its own capital constraint.
Setting \code{n\_banks = 1}, the default, recovers the \modelname{}.

A distinguishing feature of this model in the literature of macro ABMs is its ability to provide aggregate macroeconomic forecasts --- such as for GDP, inflation, and employment --- and sector-specific forecasts. 
The rich micro-structure of the model, combined with adaptive learning, allows agents to form expectations based on past behaviours and observed outcomes. 
For example, households and firms rely on simple autoregressive (AR(1)) forecasting rules. Meanwhile, decentralised market interactions are characterised by search and matching mechanisms, which introduce trade frictions and market dynamics.

\begin{figure*}[t]
\centering
\includegraphics[width=0.8\textwidth,trim={0cm 0cm 0cm 0cm},clip]{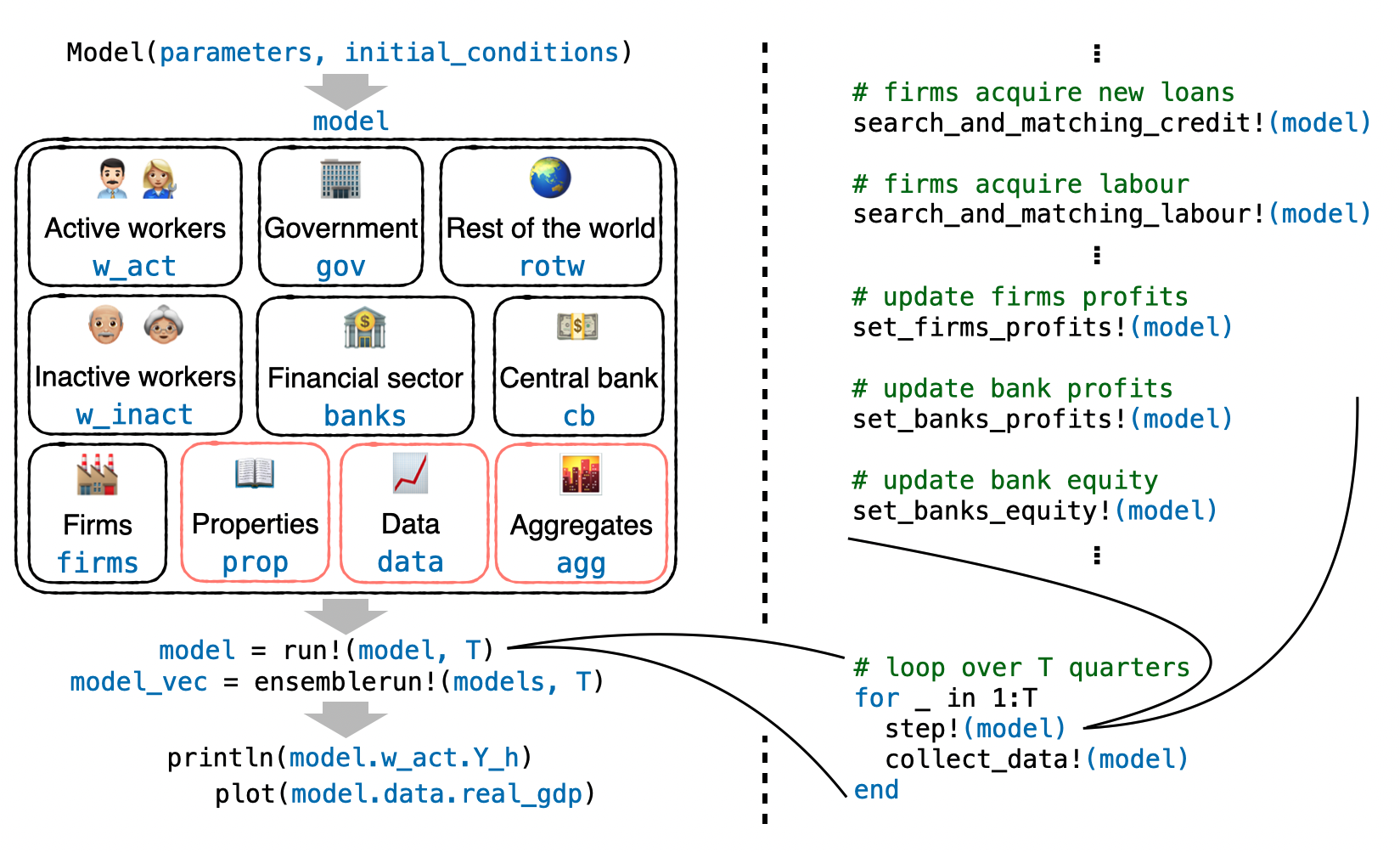}
\caption{
{\bf Illustration of the workflow and main objects of the software}. 
The constructor \code{Model} takes as input the two dictionaries \code{parameters} and \code{initial\_conditions}, returning an object named \code{model}.
This \code{model} object is composed of distinct objects representing the different agents of the economy, such as \code{gov} and \code{firms} and three other objects that do \emph{not} represent agents but store properties and aggregate variables, i.e., \code{prop}, \code{data}  and \code{agg}.
The \code{model} object serves as input for \code{run!}, which iterates over $T$ steps, while a collection of independently initialised \code{models} serves as input for \code{ensemblerun!}, which performs \code{n\_sims} independent Monte Carlo runs.
Simulation results are stored in the \code{model} object for a single run, or in the \code{model\_vec} object for multiple runs, both of which allow easy inspection and visualisation of attributes for any agent type as well as of the aggregate variables tracked during the simulation.
Similarly, we can inspect any attribute of any of the agent classes.
A simulation run simply loops over the function \code{step!}. One step consists of a series of well-defined function calls updating specific attributes of certain agent types, making up a modular and transparent simulation process.
}
\label{fig:software-structure}
\end{figure*}

\section{Software structure and main features}
\label{sec:software-structure}

\vspace{0.2cm}
\noindent \textbf{Initialisation.}
To instantiate a model, it is sufficient to use its constructor function \code{Model}.
This function takes two inputs: the dictionaries \code{parameters} and \code{initial\_conditions}. 
The two dictionaries must be defined using specific keyword strings, such as \code{"tau\_INC"} for the taxation rate on income or \code{"mu"} for the risk premium over the policy rate charged by commercial banks to issue loans.
We use the same keywords used in the original parametrisation, available from the supplementary material of \cite{poledna2023economic}.
This ensures full compatibility with the Matlab parametrisations provided by the original paper, allowing users to easily load the Matlab file and use the resulting variables for initialisation.
For quick experimentation, we include in the package the original parametrisation for Austria in the first quarter of 2010.
We also provide an experimental parametrisation for Italy.
However, we believe that the generation of parameters and initial conditions should ultimately be handled in a dedicated package specifically tailored to this purpose.

\vspace{0.2cm}
\noindent \textbf{Structure.}
The different classes of macroeconomic agents described in the previous section are implemented in \beforeit{} as separate structure of arrays (SoA) objects, which in Julia are defined via ``structs''.
%
%
Specifically, the package defines the following objects: \code{w\_act} (for households consisting of active workers), \code{w\_inact} (for households consisting of inactive workers), \code{firms} (for non-financial corporations), \code{banks} (for the financial sector, composed of one or more banks), \code{cb} (for the central bank), \code{gov} (for the government)
, \code{rotw} (for the rest of the world).
Each object contains specific and thoroughly documented attributes corresponding to the key features of the respective agent type.
For instance, the \code{w\_act} struct includes an attribute \code{Y\_h}, which stores the income of all active workers at any given simulation time-step.
The attribute names across all agent types are chosen to align closely with the equations and notation used in the original publication \cite{poledna2023economic} to facilitate a clear connection between the code and the theoretical model.
%

Finally, the entire model is encapsulated within a struct named \code{model}, which includes the seven agent groups as its attributes, along with three additional objects: the \code{prop} struct, which stores the properties of the simulation run, the \code{agg} struct, which holds the aggregate variables representing the overall state of the national economy being simulated, and the \code{data} struct, which holds the data collected during the simulation run.
The left panel of Figure~\ref{fig:software-structure} visually illustrates this hierarchical structure. 
The \code{model} struct is represented as a large rectangle containing ten smaller rectangles, corresponding to the seven agent classes and the three supplementary objects, \code{prop}, \code{data} and \code{agg}.

\vspace{0.2cm}
\noindent \textbf{Inspection.}
After a model is created as described above, its attributes can be inspected at any time using the common `dot' notation. 
For example, the income vector of all active workers can be printed using \code{println(model.w\_act.Y\_h)}, as illustrated in Figure~\ref{fig:software-structure}.
This feature can be especially convenient in interactive scripts or Jupyter notebooks to quickly verify the effects of specific code changes or interventions in the model.

\begin{figure*}[ht]
\centering
\includegraphics[width=0.9\linewidth]{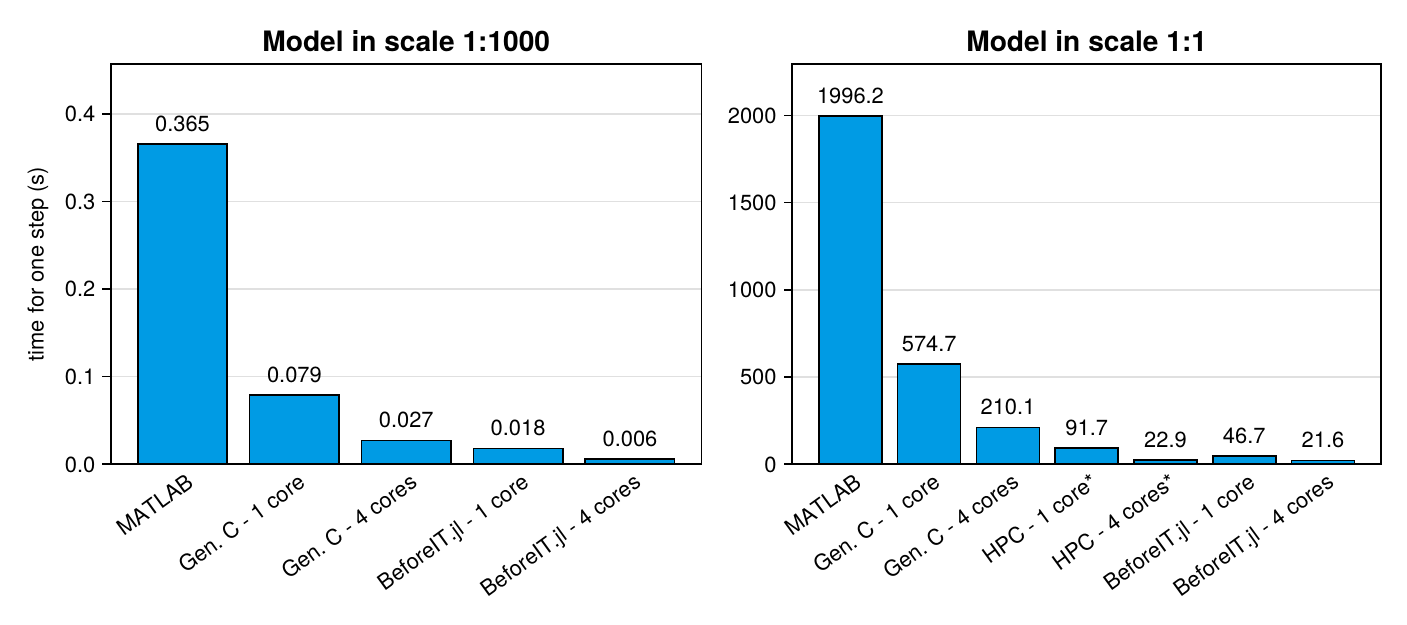}
\vspace{-10px}
\caption{{\bf A benchmark of the computational efficiency for different implementations.} 
Mean time to run one step of the model calibrated on the Austrian economy with around 8 000 agents and 8 million agents using the original Matlab code, the C code generated by the Matlab Coder toolkit, the HPC implementation and \beforeit{}. The benchmarks were executed on Linux x86\_64 with an AMD Ryzen 5 5600H CPU and 16 GB of RAM, except for the HPC results, which were only estimated from \cite{gill2021high}.
}
\label{fig:speed}
\end{figure*}

\vspace{0.2cm}
\noindent \textbf{Simulation.}
Once a model is created using the appropriate function, a simulation can be executed with a single line of code by calling the function \code{run!}.
This function takes the model object as input and returns the updated model after the simulation run.
Note that, in Julia, the exclamation mark at the end of a function denotes a function that changes its arguments in-place.
In this case, \code{run!} changes the model in-place, and hence the \code{model} object after the call stores the final state of the model.
In contrast to \code{step!}, the \code{run!} function collects data after each step. The \code{data} object has well-documented, intuitively named attributes, such as \code{nominal\_gdp} or \code{real\_household\_consumption}.
By default, 32 variables are tracked, five of which are recorded separately for each bank, but the set can be expanded by customising the data tracker.
All variables in the \code{data} object can be inspected using dot notation, for example, by calling \code{plot(data.real\_gdp)}.
It is also worth noting that the function for running a single simulation includes a keyword argument, \code{parallel}, which defaults to \code{false}.
This option allows the consumption goods market for different industrial sectors to run on separate threads, utilising as many threads as are available to the Julia session.
It is also straightforward to run a large number of Monte Carlo repetitions of the same simulation using the function \code{ensemblerun!}.
This function takes a collection of models (a vector or a generator) and the number of steps as input and returns a vector of model objects that can also be inspected using dot notation.
A keyword parameter, \code{parallel}, is available also for this function and defaults to \code{true}.
The multi-threading here is more efficient than the intra-model multi-threading across industrial sectors, as the threads operate on independent runs.

\vspace{0.2cm}
\noindent \textbf{Modularity.}
The right panel of Figure~\ref{fig:software-structure} illustrates the inner workings of the model runs.
Specifically, the figure shows that the \code{run!} function is essentially a simple for loop over a desired number of quarters, \code{T}, of the function \code{step!}, combined with a function to update the data tracker with the latest model data.
The figure also presents an excerpt from \code{step!}, highlighting how the main simulation loop consists of a sequence of very readable functions.
Each function takes the entire \code{model} as input and updates a specific set of attributes in-place, the trailing exclamation mark again denoting an in-place operation by Julia convention.
For example, \code{set\_banks\_profits!} takes \code{model} as input and updates the bank profits stored in the attribute \code{model.banks.Pi\_k}.
This clear modular structure allows for precise changes and extensions to the model by simply redefining specific functions.

\begin{figure*}[ht]
\begin{minipage}[b]{0.4\linewidth}
\centering
    \begin{jllisting}[language=julia, style=mystyle]
    import BeforeIT as Bit
    import Plots, StatsPlots

    # load parameters and initial conditions
    p = Bit.AUSTRIA2010Q1.parameters
    ic = Bit.AUSTRIA2010Q1.initial_conditions

    # define a simulation length
    T = 20

    # initialise 8 model objects
    models = (Bit.Model(p, ic) for _ in 1:8)

    # run 8 simulations in parallel
    model_vector = Bit.ensemblerun!(models, T)

    # plot the results
    plots = Bit.plot_data_vector(model_vector)
    Plots.plot(plots...)
    \end{jllisting}
\end{minipage}
\hspace{0.2cm}
\begin{minipage}[b]{0.6\linewidth}
\centering
\includegraphics[width=\textwidth]{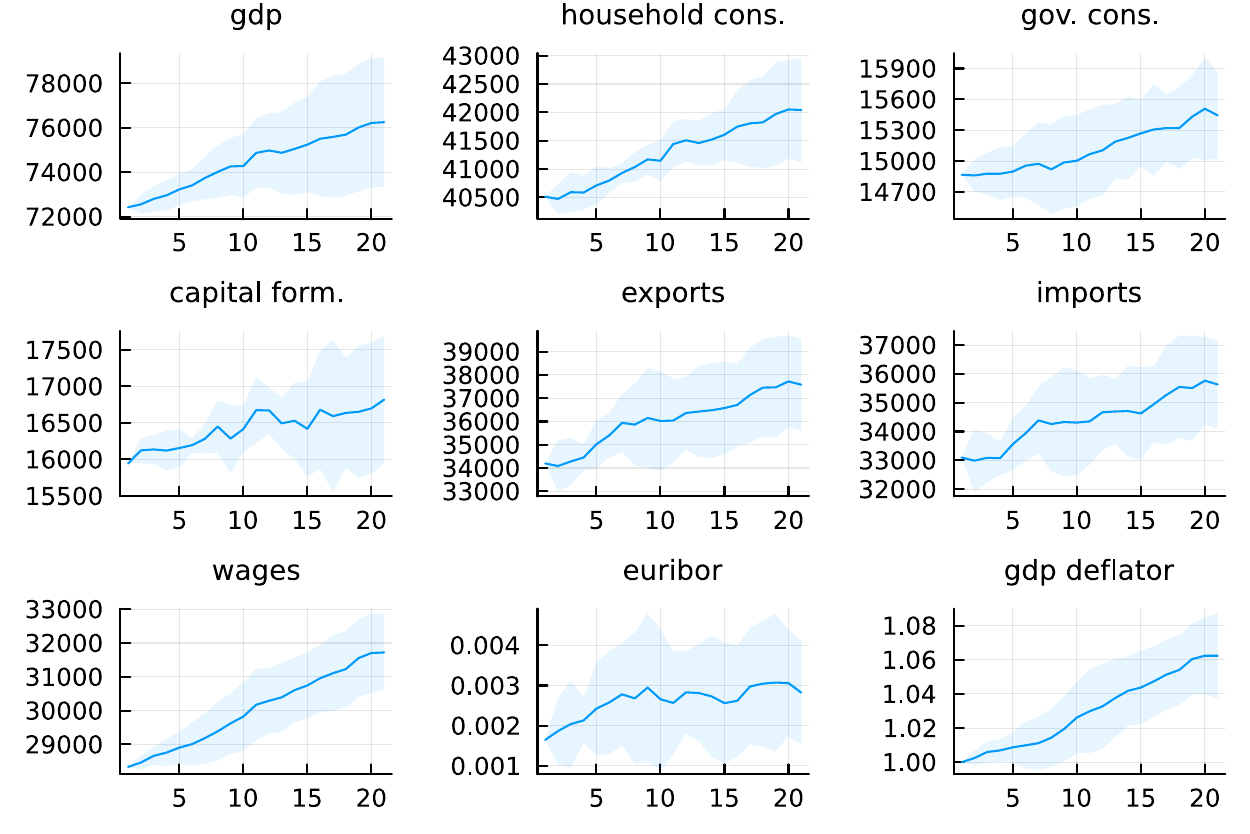}
\end{minipage}
\vspace{-10px}
\caption{{\bf A simple script and its resulting graphs.}
\label{fig:simple_script}
The left panel shows a minimal script to run the model.
We import the package, load the original parametrisation for Austria in the first quarter of 2010, and specify a given number of quarters to simulate.
Then we call the \code{Model} function eight times, one for each Monte Carlo run, and finally the \code{ensemblerun!} function.
Illustrative charts of some variables within \code{model\_vector} are shown in the right panel. Solid lines are averages across the eight simulations, while semi-transparent regions span one standard deviation.
}
\end{figure*}

\vspace{0.2cm}
\noindent \textbf{Speed.}
In addition to being easy to use and extend, \beforeit{} is highly computationally efficient. 
In Figure~\ref{fig:speed} we show the efficiency of the software by benchmarking the mean time required to run the model of Austria at scale 1:1000, hence simulating around 8000 agents, and the model at scale 1:1, with around 8 million agents, for one step against the Matlab implementation, the C implementation generated with the Matlab-Coder toolbox and the HPC model developed in \cite{gill2021high}.
Our tests show that \beforeit{} is approximately 20 times faster than the Matlab implementation and 4 times faster than the Matlab-generated C code for the small version in the single-threaded case and, respectively, 43 times and 12 times faster when 8 million agents are simulated. 
Our simulation appears to have better run-time performance compared to the HPC implementation of the model by looking at the running times reported in the original article \cite{gill2021high}. 
However, given the unavailability of the source code and the different execution environments, we cannot measure the exact speed-up.
The efficiency of \beforeit{} is largely due to Julia's just-in-time compilation, which enables a performance comparable to C while making the code significantly easier to develop, maintain and understand \cite{sells2020julia,hunold2020benchmarking,lin2021comparing}.
Additional performance gains are achieved by employing more optimised algorithms than those used in the original Matlab version. 
The greatest part of the performance improvement was made possible by using dynamic weighted sampling techniques \cite{slepoy2008,d2022dynamic}. In particular, we used the state-of-the-art implementation provided in \cite{hafner2025sampling}, which allowed to simulate the market exchange process for goods much more efficiently.
Notably, even without the use of such advanced algorithms, our software proved to be approximately on par with the C implementation.
The figure also illustrates that applying multi-threading across different industrial sectors can further accelerate the simulation, albeit the speed-up is sub-linear relative to the number of cores employed; even if we believe this aspect could be improved in the future, the scalability of the HPC implementation is still better than what BeforeIT.jl currently achieves.
However, when parallelising over different independent Monte Carlo runs, BeforeIT.jl offers the best performance, without sacrificing ease of use.

\noindent \textbf{Testing.}
\beforeit{} includes a large and expanding suite of \emph{unit tests} to ensure its quality and reliability.
The current set of unit tests can be divided into three classes.
At the lowest level, we test specific functions representing individual agent actions against the expected behaviour.
As an example, we test that the function \code{taylor\_rule} with specific values of input arguments returns the same number that one can compute with pen and paper.
At the aggregate level, we test that the code respects several accounting identities. 
For instance, we check that the national income identity holds after every step, i.e., that the GDP equals the sum of all expenditures.
Finally, at the model level, we test that a deterministic version of the model exactly matches the detailed behaviour of a deterministic version of the original Matlab code.
Additionally, stochastic versions of the two models are rigorously tested for coherence through statistical analysis.

These layers of tests are run automatically through GitHub's continuous integration workflows every time the code is modified to collectively ensure both the accuracy and robustness of the software.

\vspace{0.2cm}
\noindent \textbf{Documentation.}
\beforeit{} is thoroughly documented.
The documentation for the main functions governing the inner workings of the package and of the agents in the model is embedded in the source code and also forms an API reference available at \url{https://bancaditalia.github.io/BeforeIT.jl}.
The webpage provided also includes several tutorials to quickly learn how to use the package with a hands-on approach.
The source code reported in the online tutorials is available in the ``examples'' folder of the GitHub repository along with other example usages.

\section{Usage illustration}
\label{sec:usage-illustration}

Here, we present some illustrative examples of how to use the package.
We provide self-contained scripts directly here where possible, but we highlight that the complete, runnable source code for all the examples discussed is available in the \code{examples} folder of the repository\footnote{\href{https://github.com/bancaditalia/BeforeIT.jl/tree/main/examples}{https://github.com/bancaditalia/BeforeIT.jl/tree/main/examples}}.

\vspace{0.2cm}
\noindent \textbf{Essential usage.}
Figure~\ref{fig:simple_script} provides a step-by-step example of typical usage of the package.
The left panel outlines the exact commands required to load the package, initialise a model, run different Monte Carlo runs of the model, and then inspect the results. 
The right panel presents a few plots of the simulation runs, with the shaded area indicating the standard error on the mean of the Monte Carlo repetitions.

\begin{figure}[!ht]
    \begin{minipage}[b]{0.99\linewidth}
    \centering
    \begin{jllisting}[language=julia, style=mystyle]
    # define a custom shock
    struct ConsumptionShock <: Bit.AbstractShock
        multiplier::Float64
        final_time::Int
    end

    # define the shock behaviour
    function (s::ConsumptionShock)(model)
        if model.agg.t == 1
            model.prop.psi *= s.multiplier
        elseif model.agg.t == s.final_time
            model.prop.psi /= s.multiplier
        end
    end

    # load parameters and initial conditions
    p = Bit.AUSTRIA2010Q1.parameters
    ic = Bit.AUSTRIA2010Q1.initial_conditions

    # set shock parameters
    my_shock = ConsumptionShock(1.02, 4) 

    # initialise models
    models = (Bit.Model(p, ic) for _ in 1:64)

    # run the shocked models
    T = 20
    model_vec = Bit.ensemblerun!(models, T; shock! = 
                                            my_shock)
    \end{jllisting}
    \end{minipage}
    \centering
    \includegraphics[width=0.4\textwidth]{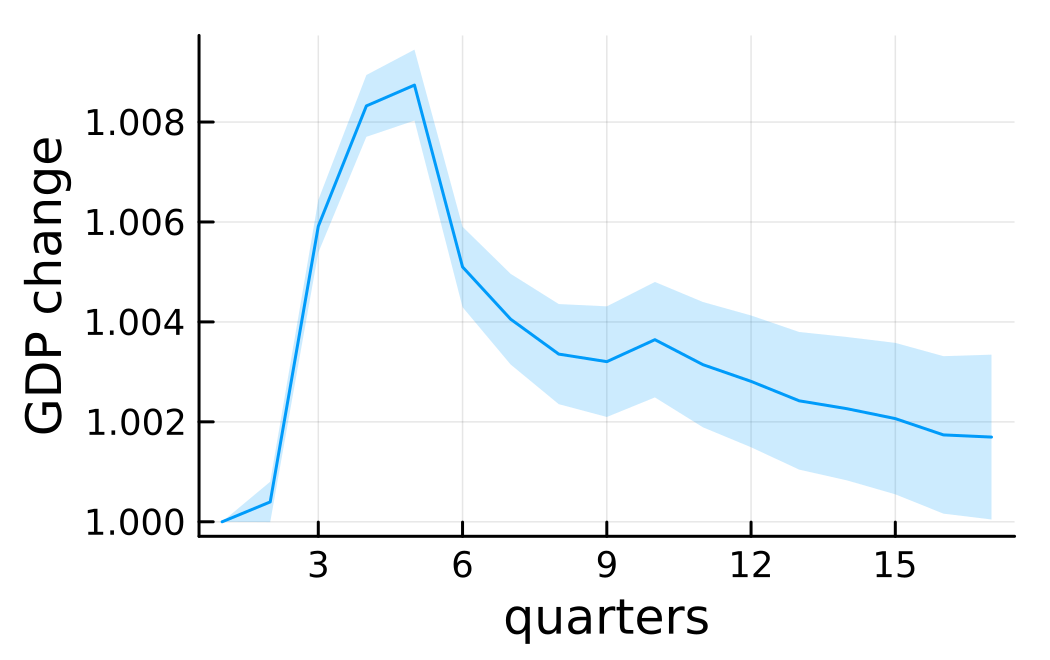}
\vspace{-10px}
\caption{{\bf A script for shocked simulations and its resulting graph.}
The top panel illustrates the usage of the interface to run simulations with shocks.
One simply defines a callable \code{struct} and passes it to the function \code{ensemblerun!} as the \code{shock!} keyword argument.
The effect of the simulated consumption shock on the GDP (ratio of shocked over baseline) is presented in the bottom panel. Solid lines are averages across the 64 simulations, while semi-transparent regions span the standard error.
}
\label{fig:shock}
\end{figure}

\begin{figure}[t]
    \centering
    \begin{minipage}[t]{0.99\linewidth}
    \begin{jllisting}[language=julia, style=mystyle]
    # define a new model object
    Bit.@object struct NewModel(Bit.Model) 
                            <: Bit.AbstractModel end

    # define a new central bank object    
    Bit.@object mutable struct NewCB(Bit.CentralBank) 
                            <: Bit.AbstractCentralBank
        avg_r::Float64
    end

    # change the default central bank behaviour
    function Bit.set_central_bank_rate!(model::
                                        NewModel)
        model.cb.r_bar = model.cb.avg_r + 0.1*randn()
    end
    
    # initialise all basic model components 
    prop           = Bit.Properties(p, ic)
    firms          = Bit.Firms(p, ic)
    w_act, w_inact = Bit.Workers(p, ic)
    banks          = Bit.Banks(p, ic)
    gov            = Bit.Government(p, ic)
    rotw           = Bit.RestOfTheWorld(p, ic)
    agg            = Bit.Aggregates(p, ic)
    data           = Bit.Data()

    # initialise the custom central bank
    cb = Bit.CentralBank(p, ic)
    newcb = NewCB(Bit.fields(cb)..., 0.02)
    
    # initialise the model
    model = NewModel((w_act, w_inact, firms, banks, 
                newcb, gov, rotw, agg, prop, data))

    # run 8 copies of the extended model
    T = 20
    models = (deepcopy(model) for _ in 1:8)
    data\_vec = Bit.ensemblerun!(models, T)
    \end{jllisting}
    \end{minipage}
\caption{{\bf A script for model extensions.}
The script illustrates the usage of macros and methods' specialisation to extend the model without needing to copy-paste code.
One can create a new \code{mutable struct} for a custom central bank as a subtype of \code{AbstractCentralBank} and inherit all standard attributes using the \code{@object} macro.
Then, one can modify specific behaviours by specialising some functions on the new model type \code{model::NewModel}, while the other functions remain unchanged.
}
\label{fig:extension}
\end{figure}

\vspace{0.2cm}
\noindent \textbf{Shocked simulations.}
We also provide a high-level interface for simulating arbitrary economic shocks to the model. 
This functionality is achieved through a ``shock'' object, implemented as a callable struct.

At the beginning of every step, and before agents' actions, the shock object is invoked.
The shock object receives the entire model as input, enabling it to modify any model attribute in-place in a highly flexible manner. 
Figure~\ref{fig:shock} illustrates this process with an example of a consumption shock.
In the figure, we first create a struct with two attributes.
Then we make the struct ``callable'' by defining a function on top of the struct. 
The function increases the propensity to consume \code{psi} by the factor \code{multiplier} until the simulation time \code{model.agg.t} reaches \code{final\_time}, when it is set back to the original level.
Finally, a specific shock (\code{my\_shock}) is instantiated with a 2\% increase in the propensity to consume for 4 quarters.
This shock is then passed as a keyword argument to the function \code{ensemblerun}\code{!}.
By default, the keyword argument is set to a dummy ``NoShock'' struct that performs no actions.
This abstraction allows for arbitrary complex shocks to be implemented straightforwardly since both the state of the model and any of its parameters are accessible as attributes of the \code{model} object and can be easily manipulated. 

\vspace{0.2cm}
\noindent \textbf{Calibration to EU nations.}
In a separate project called CalibrateBeforeIT.jl \cite{calibratebeforeitjl} we provide functions and exemplary scripts which allow the user to generate the necessary parameters and initial conditions for other EU nations.

The code in the CalibrateBeforeIT.jl package is based on the Matlab and PostgreSQL
scripts used in \cite{poledna2023economic} (which were already extended to work 
for all EU27 countries at that point) and were kindly shared with us. To ease the
accessibility of this work, we translated the scripts to Julia, updated Eurostat 
sources and made the process more robust against missing data. The only 
prerequisite for this repository is Julia and there are no outside technical 
dependencies.

We showcase the usage of the package in three scripts (available in the repository). 
First, usage of the helper functions to a) download data from a data snapshot at Zenodo\footnote{\href{https://doi.org/10.5281/zenodo.18698434}{https://doi.org/10.5281/zenodo.18698434}} to be able
to replicate our results or b) download the latest available data from Eurostat. 
The second script shows the application of some necessary data transformations (e.g.\ appending two data sources to have longer timeseries available). Finally, a third script shows the helper functions that are used to query the data sources for the timeseries necessary to construct the initial conditions and parameters. 
At our Zenodo repository, we also provide the resulting country-specific calibration objects (which can be used to calculate initial conditions for the \beforeit{} model).

At the moment of writing, the CalibrateBeforeIT.jl package 
works for all EU nations for all quarters from 2015 to 2023.\footnote{Applying the calibration process for years before 2015 and after 2023 is possible, but more prone to fail due to missing data.}

For quick experimentation, users of \beforeit{} can use the \code{download\_zenodo\_calibration\_object} function to directly download the calibration object of any of the 27 EU member states and  calculate \code{parameters} and \code{initial\_conditions}
for a given quarter (specified by the last day of that 
quarter).
A practical example is provided in Figure~\ref{fig:zenodo_calibration}.
\begin{figure}[t]
\begin{minipage}[h]{0.99\linewidth}
    \begin{jllisting}[language=julia, style=mystyle]
    using Dates

    # download the calibration object for France
    co = Bit.download_zenodo_calibration_object("FR")

    # calculate initial state for a given time
    calibration_date = DateTime(2022, 12, 31)
    parameters, initial_conditions = 
       Bit.get_params_and_initial_conditions(
          co, calibration_date; scale = 1 / 1000)
    \end{jllisting}
\end{minipage}
\caption{{\bf Downloading a country calibration object.} The \code{download\_zenodo\_calibration\_object} function retrieves the pre-computed calibration object for any of the 27 EU member states from the Zenodo data snapshot, here France (\code{"FR"}). The object is then passed to \code{get\_params\_and\_initial\_conditions} together with a calibration date (the last day of the target quarter) to obtain the \code{parameters} and \code{initial\_conditions} dictionaries needed to instantiate a \code{Model}. The \code{scale} keyword sets the agent-to-population ratio, here $1{:}1000$.}
\label{fig:zenodo_calibration}
\end{figure}

We are actively developing the CalibrateBeforeIt.jl package 
to extend its country  and time coverage and remedy 
any existing inconsistencies.

\begin{figure*}[t]
    \centering
    \includegraphics[width=1\linewidth]{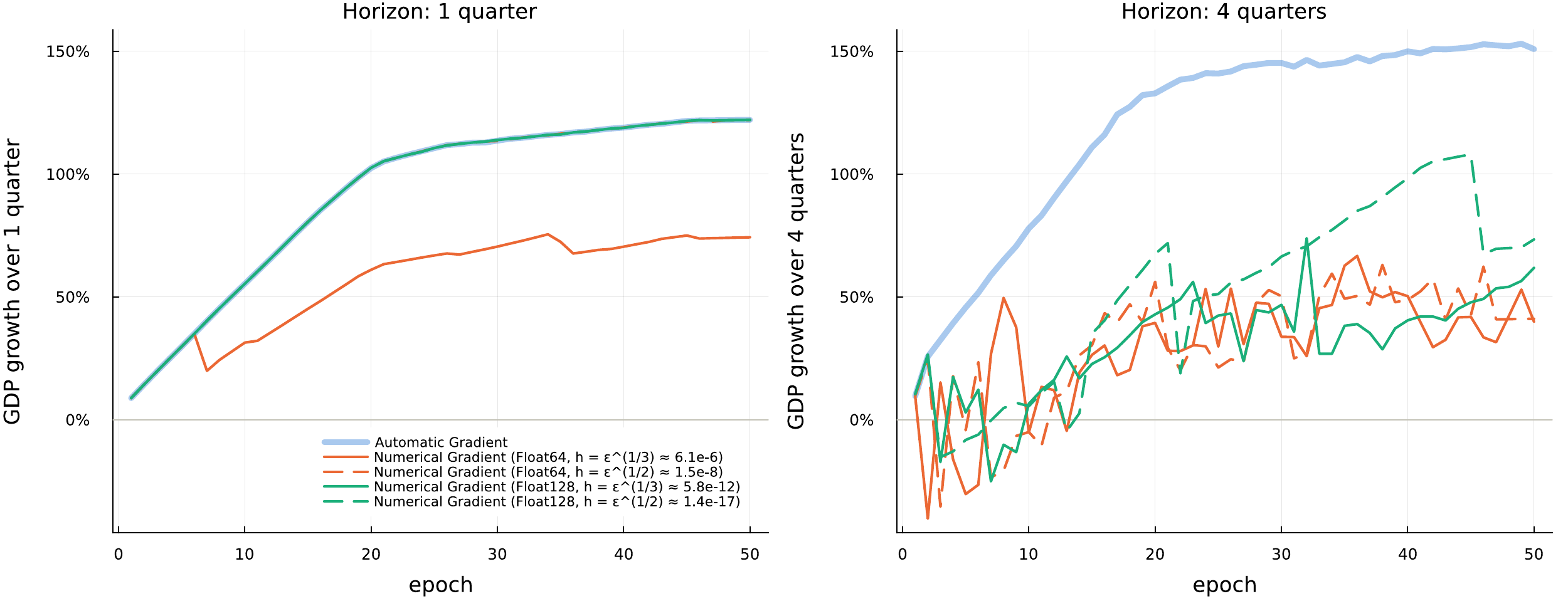}
    \caption{{\bf GDP optimisation trajectories by gradient method.} Mean percentage GDP growth over 10 held-out random seeds along the optimisation, for an objective with horizons of one quarter (left) and four quarters (right). The Mooncake pathwise automatic gradient is compared against central finite differences computed in 64-bit and 128-bit floating-point precision, each with the relative steps $h = \varepsilon^{1/3}$ and $h = \varepsilon^{1/2}$. Over one quarter, trajectories corresponding to 64-bit and 128-bit numerical gradients with $h = \varepsilon^{1/2}$ overlap with the trajectory corresponding to automatic gradient, while over four quarters trajectories corresponding to every numerical gradient starkly depart from it.}
    \label{fig:ad_convergence}
\end{figure*}

\vspace{0.2cm}
\noindent \textbf{Automatic differentiation.}
Automatic differentiation (AD) is a family of techniques for computing derivatives of functions implemented as computer programs by systematically applying the chain rule to the elementary operations performed during execution. 
In machine learning, AD is widely used for training deep neural networks \cite{rumelhart1986learning, baydin2018automatic}. 
More recently, AD and differentiable simulation have been explored for agent-based models, where gradients of simulator outputs with respect to parameters can accelerate tasks such as calibration, optimisation, and sensitivity analysis, provided the model is made suitably differentiable ~\cite{andelfinger2021differentiable, kotthoff2022calibrating, dyer2023gradient}.

Among the many AD tools available in Julia \cite{revels2016autodiff,innes2019autodiff,moses2020autodiff, moses2021autodiff,arya2022autodiff,tebbutt2024autodiff}, \beforeit{} currently leverages the Mooncake compiler \cite{tebbutt2024autodiff} to allow for automatic differentiation.
%
Achieving this compatibility required refactoring specific segments of the codebase to circumvent non-differentiable code blocks.
For instance, because Mooncake cannot differentiate through foreign function calls (such as LAPACK routines), the singular value decomposition (SVD) used to estimate expected GDP and inflation had to be re-implemented in pure Julia. 
While currently unsupported, future integration with Enzyme~\cite{moses2021autodiff} would be beneficial for cross-validating gradient computations across distinct AD backends.

To verify the effectiveness of using automatic differentiation, we run a simple dummy optimisation on the GDP of the Austrian model calibrated to 2010Q1.
Specifically, we maximise the objective
$$\frac{\text{GDP}(\theta)_{t=H} -  \text{GDP}(\theta)_{t=0}}{\text{GDP}(\theta)_{t=0}}  - \left(\frac{\text{GDP}(\theta)_{t=0} - \text{GDP}(\theta_0)_{t=0}}{\text{GDP}(\theta_0)_{t=0}}\right)^2,$$
where $\text{GDP}(\theta_0)_{t=0}$ represents the initial GDP of the original Austrian parametrisation $\theta_0$, while $\text{GDP}(\theta)_{t=0}$ and $\text{GDP}(\theta)_{t=H}$ denote the initial GDP of a new parametrisation $\theta$ and its GDP after a horizon of $H$ quarters, respectively.
This optimisation tries to anchor the initial GDP to the same level of the original parametrisation and, at the same time, induce a greater GDP growth over the following $H$ quarters.
To make the gradient optimisation more stable, only a subset of 17 core policy and behavioural parameters (which include fiscal rates, household behaviour parameters and banking/financial parameters) have been allowed to vary in the optimisation, each within $[0, 1]$.

We optimise the objective by stochastic gradient ascent over 50 epochs, computing each gradient on a new random seed, so that the optimisation targets the expected GDP growth rather than a single random realisation of it.
We compare the automatic gradient extracted with Mooncake.jl against numerical gradients computed by central finite differences in 64-bit and 128-bit floating-point precision, made possible by \beforeit{} supporting any floating-point type.
Within one numerical gradient, all evaluations of the objective share the same random seed (common random numbers), and for each precision we use two textbook choices for relative steps: $h = \varepsilon^{1/3}$ and $h = \varepsilon^{1/2}$, where $\varepsilon$ is the machine precision \cite{dennis1996numerical, nocedal2006numerical}.
The parameters reached at each epoch are scored by the GDP growth they produce on 10 held-out random seeds that no gradient uses.

Figure~\ref{fig:ad_convergence} shows the results for horizons of one and four quarters.
Over one quarter, both the 64-bit and 128-bit numerical gradients with the smaller step track the automatic gradient closely, within half a percentage point at every epoch, and reaching a GDP growth of about 122\%, whereas the 64-bit numerical gradient with the larger step departs from the the automatic gradient after a few epochs, resulting in a GDP growth of about 74\%.
Over four quarters, instead, the automatic gradient steadily raises GDP growth to about 150\%, while every numerical gradient, including the 128-bit ones, leads to an erratic optimisation ultimately yielding a GDP growth between 40\% and 73\%.
This is because a model step is only piecewise smooth: firms hire whole workers and buyers are matched with sellers at random, so a small change in a parameter can flip a discrete outcome and make the objective jump; a finite-difference step that straddles such a jump returns the size of the jump divided by $2h$ rather than a derivative.
Each additional quarter adds further discrete outcomes and, from the second quarter onwards, the re-estimated expectations amplify rounding errors, so the range of usable steps shrinks from both sides as the horizon grows: over four quarters no 64-bit step is usable, and even 128-bit steps are not reliable throughout the optimisation.
The automatic gradient, which differentiates the realised execution path, is not affected by these jumps.
Note, however, that for the same reason the automatic gradient computed with Mooncake is not exact: by following only the realised execution trace, it ignores how the jumps contribute to the change in the expected objective.
%
%
When a model contains discrete randomness (as in our search-and-matching markets), the resulting gradient can be biased \cite{querabofarull2025automatic}. 
In practice, the results in Figure~\ref{fig:ad_convergence} suggest that any such bias is small for our model: over one quarter, the automatic gradient follows the same optimisation path as the reliable numerical gradients and, over four quarters, it keeps improving GDP growth on the held-out seeds.

\begin{figure}[t]
    \centering
    \includegraphics[width=1\linewidth]{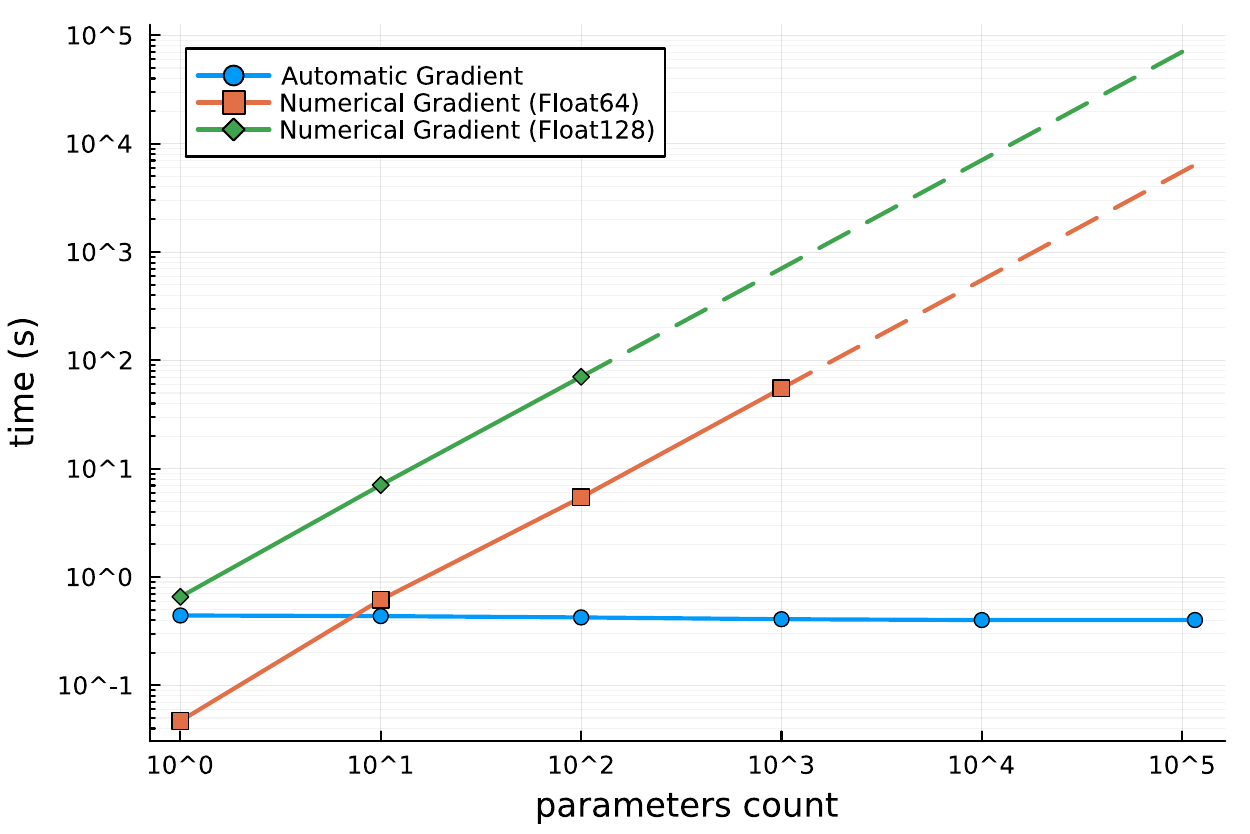}
    \caption{{\bf Performance of gradient computation.} Wall-clock time to compute the gradient of the objective with respect to an increasing subset of the model parameters, comparing the Mooncake automatic gradient against central finite differences.}
    \label{fig:ad_performance}
\end{figure}


Performance-wise, automatic differentiation changes the scaling of the gradient computation dramatically. 
%
%
As shown in Figure~\ref{fig:ad_performance}, the runtime of the automatic gradient remains essentially flat as the number of active parameters grows from one to the full model parametrisation, at roughly the cost of differentiating one full model step. 
Instead, central finite-difference gradients scale linearly with the number of parameters,
taking about one minute for 1 000 parameters, while the automatic gradient remains practical for the full 106 955-parameter state. 
Thus, automatic differentiation is not only more convenient, but also makes high-dimensional gradient computations feasible.

\vspace{0.2cm}
\noindent \textbf{Extensions.}
We include in \beforeit{} specific macros that allow for easily extending the original agent classes without copying and pasting the entire codebase.
In Figure~\ref{fig:extension} we provide an example of how to leverage these macros.
Suppose we want to extend the model by introducing a new type of central bank with custom behaviour.
We do this by creating a custom struct called \code{NewCB}.
Crucially, we define this to be a subtype of \code{AbstractCentralBank}.
This allows the new struct to seamlessly inherit all existing functions defined in \beforeit{} for central banks.

In our new struct, we want to inherit all existing attributes of the standard central banks in the package and add additional attributes.
We do this using the macro \code{@object}, which essentially copy-pastes all the attributes of the inherited object, which here is \code{CentralBank}, within the newly created one.
Additional attributes can then be defined as needed.
Finally, we customise the behaviour of the central bank by defining a new \code{set\_central\_bank\_rate!} function.
In this function, we specify that the \code{model} argument is of the custom type \code{NewModel}.
In this way, through Julia's powerful multiple-dispatch mechanism, the compiled code will use this function instead of the default one, and will instead use the default functions for any other behaviours of the central bank as provided in the package implementation.

%

%


\section{Systematic model improvements}                     \label{sec:model_improvements}

The extension mechanism introduced in Section~\ref{sec:usage-illustration} enables more than ad-hoc experimentation: it provides a principled framework for \emph{systematic} model improvement. By pairing model variants with an automated multicountry forecasting pipeline, researchers can rigorously evaluate alternative   
modelling assumptions across all calibrated economies using standardised error metrics. This section describes the two main ingredients: (i) how full model variants are defined within \beforeit{}, and (ii) how they are
evaluated at scale, illustrating the workflow with a concrete example.

\vspace{0.2cm}
\noindent \textbf{Defining model variants.}
A model variant in \beforeit{} is a self-contained Julia file that defines new agent types and overrides specific behavioural functions via multiple dispatch, leaving the rest of the simulation unchanged. The \code{@object} macro copies all fields from a parent type and appends any additional ones, while the Julia type system ensures that only the overridden functions are replaced. Two representative variants are shipped with the package.

The first, \code{ModelCANVAS}, implements the demand-pull pricing mechanism of \cite{hommes2025canvas} by overriding three functions: \code{firms\_expectations\_and\_decisions}, which introduces firm-level quantity and price adjustments based on excess demand; \code{central\_bank\_rate}, which replaces the fixed-parameter Taylor rule with an adaptive version that re-estimates its coefficients each step; and \code{growth\_inflation\_EA}, which records the euro-area time series needed for the adaptive estimation. These overrides require two extended agent types: \code{CentralBankCANVAS}, which stores the historical interest rate series, and \code{RestOfTheWorldCANVAS}, which stores euro-area GDP and inflation histories. All other agents (firms, households, banks, government) remain unchanged.
We also include two further backwards-compatible improvements to \code{ModelCANVAS} that allow trajectories to converge over much longer time horizons (around 100 steps).
First, we implement Calvo pricing: the parameter \code{theta\_calvo} allows to specify the fraction of firms that can update their prices in each step.
Second, we use an exponential moving average with smoothing factor \code{omega\_pi} to compute inflation expectations.
\code{ModelCANVAS} falls back to the behaviour in \cite{hommes2025canvas} by setting \code{theta\_calvo} $ = 1$ and \code{omega\_pi} $ = 0$, while typical values for converge over longer time horizons are \code{theta\_calvo} $ = 0.25$ and \code{omega\_pi} $ = 0.5$.

The second variant, \code{ModelGR}, replaces the log-level AR(1) processes used for expectations and exogenous drivers with growth-rate AR(1) specifications,

\begin{equation}
  g_t = \alpha \, g_{t-1} + \beta + \varepsilon_t, \qquad X_t = X_{t-1} \cdot (1 + g_t),
\end{equation}
where $g_t$ denotes the growth rate of the variable $X_t$. This affects GDP expectations, inflation expectations, government consumption, and trade volumes, and requires two new agent types, \code{GovernmentGR} and \code{RestOfTheWorldGR}, that store the estimated AR(1) coefficients and lagged growth rates. In both cases, the core simulation loop (\code{step!}) remains completely untouched.
Julia's compiler automatically dispatches to the overridden methods based on the model's concrete type, ensuring that every variant is evaluated under identical market-clearing, accounting, and aggregation rules and thereby isolating the effect of the modelling change.

\begin{figure}[ht!]
\centering
\begin{minipage}{0.50\textwidth}
\begin{jllisting}[language=julia, style=mystyle, framexleftmargin=1pt]
import BeforeIT as Bit
using Dates

# countries to process (2-letter codes)
countries = ["IT", "AT", "ES"]
quarters = DateTime(2010,3,31):Month(3):DateTime(2019,12,31)
T, n_sims = 12, 4

# pick the variant; the variant name (and hence the
# output folders) is derived from it, so they agree
model_constructor = Bit.ModelGR
model_variant     = string(nameof(model_constructor))

# --- Simulation phase ---
for country in countries
    cal = Bit.download_zenodo_calibration_object(country)
    folder = "data/$country"
    Bit.save_all_params_and_initial_conditions(
        cal, folder; scale = 0.0005,
        first_calibration_date = first(quarters),
        last_calibration_date  = last(quarters),
    )
    Bit.save_all_simulations(folder; T, n_sims,
        model_constructor,
        simulation_folder = "simulations/$model_variant")
    Bit.save_all_predictions_from_sims(folder, cal.data;
        simulation_suffix = "simulations/$model_variant",
        prediction_suffix = "abm_predictions/$model_variant")
end

# --- Analysis phase ---
for country in countries
    cal = Bit.download_zenodo_calibration_object(country)
    Bit.error_table_ar(country, cal.ea, cal.data,
        quarters, Bit.forecast_horizons; model_variant)
    Bit.error_table_abm(country, cal.ea, cal.data,
        quarters, Bit.forecast_horizons; model_variant,
        prediction_folder = "abm_predictions/$model_variant")
end
\end{jllisting}
\end{minipage}
 \caption{{\bf The multicountry forecasting pipeline.}
 The same script drives the two phases of the pipeline. 
 In the \emph{simulation phase}, the chosen model variant (here \code{ModelGR}) is calibrated to each country (downloading calibration objects from Zenodo) and run. Forecasted trajectories are saved to disk. 
 In the \emph{analysis phase}, saved forecasts are compared to an AR benchmark using standardised error metrics. 
 One can select a different model variant via \code{model\_constructor}; the output folders automatically reflect the model variant.}
\label{fig:pipeline}
\end{figure}

\begin{figure}[ht!]
    \centering
        \includegraphics[width=1.0\linewidth, trim={2cm 2cm 2cm 2cm},clip]{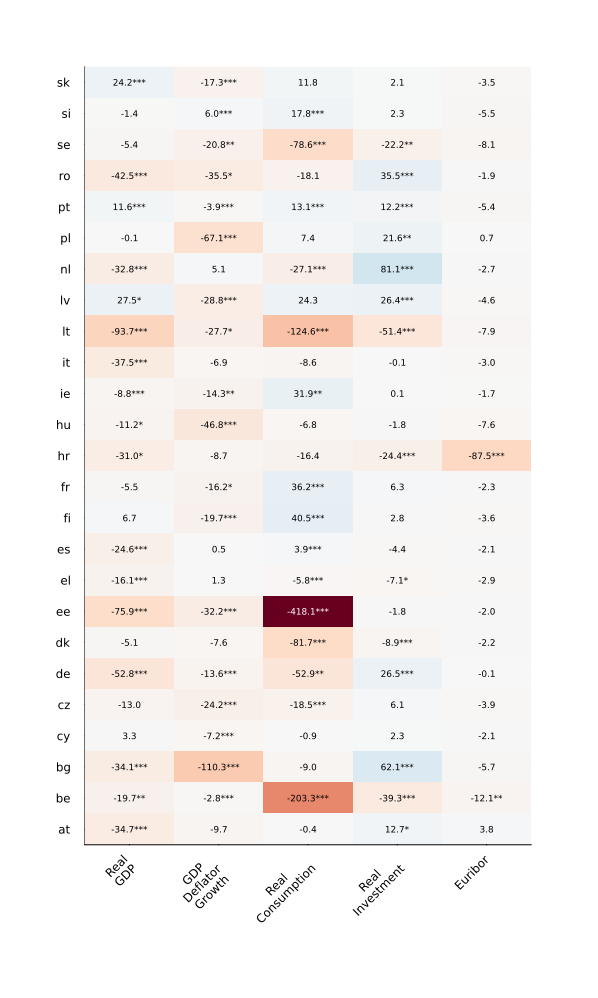}
        \vspace{-1cm}
        \caption{{\bf RMSE improvement of \code{ModelGR} over a univariate AR benchmark.}
            Each cell shows $(\text{RMSE}_\text{AR} - \text{RMSE}_\text{ABM}) \times 100$ over a four-quarter horizon, averaged over all quarters from 2010Q1 to 2019Q4. Blue cells indicate that the ABM produces lower forecast errors than the AR; red cells indicate the opposite. Stars denote significance levels of the Diebold--Mariano, whose null hypothesis is that the two models have equal predictive accuracy ($^{*}p<0.1$, $^{**}p<0.05$, $^{***}p<0.01$).}
    \label{fig:rmse_heatmap}
\end{figure}

\vspace{0.2cm}
\noindent \textbf{Multicountry forecasting pipeline.}
To evaluate model variants at scale, \beforeit{} includes an automated pipeline that runs out-of-sample forecasting exercises across a list of calibrated countries.
A variant is identified by a model constructor, from which the variant name is derived; output is automatically organised under \code{data/\{country\}/analysis/\{variant\}/}, so adding a new variant requires no changes to the pipeline code itself.

The pipeline has two phases, both driven by the same script, shown in Figure~\ref{fig:pipeline}.

In the \emph{simulation phase}, three helper functions are called in sequence for each country. First, \code{save\_all\_params\_and\_initial\_conditions} iterates over all starting quarters in the chosen range (here 2010Q1 to 2019Q4), generating and saving to disk the \code{parameters} and \code{initial\_conditions} for each quarter. Then, \code{save\_all\_simulations} loads each of these, instantiates the model using the provided \code{model\_constructor}, runs \code{n\_sims} Monte Carlo simulations of length \code{T}, and saves the resulting ensembles. Finally, \code{save\_all\_predictions\_from\_sims} aligns the saved simulations with the real data to produce the forecast trajectories. When \code{model\_constructor} is omitted, the standard \code{Model} constructor is used, corresponding to the base model.

In the \emph{analysis phase}, forecast accuracy is assessed using the root mean squared error (RMSE) and bias at horizons of 1, 2, 4, 8, and 12 quarters for five key macroeconomic variables: real GDP, GDP deflator growth, real household consumption, real investment, and the Euribor rate. Statistical significance of the bias is tested using a $t$-test with Newey--West standard errors to account for serial correlation
in forecast errors. Model comparisons are performed via the Diebold--Mariano test, which assesses whether the difference in squared forecast errors between two models is statistically significant. Results are stored per country in \code{data/\{country\}/analysis/\{variant\}/} and then aggregated across countries to produce
summary tables and heatmaps.

\vspace{0.2cm}
\noindent \textbf{Example results.}
Figure~\ref{fig:rmse_heatmap} illustrates the kind of output the pipeline produces, showing the RMSE difference between \code{ModelGR} and a univariate AR benchmark across a large number of calibrated EU countries. 
Each cell reports $(\text{RMSE}_\text{AR} - \text{RMSE}_\text{ABM}) \times 100$, so that positive values (blue) indicate that the ABM outperforms the AR and negative values (red) indicate the reverse; significance stars denote rejection of equal predictive accuracy under the Diebold--Mariano test.

The results are heterogeneous across countries and variables.
The ABM achieves notably better forecast accuracy for real investment in several countries (e.g., the Netherlands and Bulgaria), while the AR benchmark tends to be more competitive for real GDP and inflation in many economies. 
Outliers such as Estonia's real consumption reflect episodes of large structural change that are difficult to capture with any fixed specification. 
This heterogeneity illustrates the value of running the pipeline at scale: patterns that would be invisible in a single-country study become apparent when results are aggregated.

\section{Conclusions}
\label{sec:conclusions}

This paper introduces \beforeit{}, a software package to build state-of-the-art macroeconomic agent-based models (macro ABMs) that are based on the recently introduced \modelname{} \cite{poledna2023economic}, in a fast, user-friendly, and extensible framework. 
By leveraging the computational efficiency and flexibility of the Julia programming language, \beforeit{} delivers high-performance code that is greatly superior to existing implementations of the \modelname{}, such as Matlab or Matlab-generated C code, while maintaining simplicity for users and developers.
Its modular structure, thorough documentation, and extensive testing ensure that the package can be easily picked up and used for applications ranging from policy analysis and economic forecasting to any model extension.
As the first open-source package to build models like \cite{poledna2023economic}, \beforeit{} fills an important gap in making the latest macro ABMs more accessible.
We believe the package can greatly facilitate experimentation, e.g., involving calibration using large volumes of data, or extensive sensitivity analyses and forecasting. 
The existing open-source Julia ecosystem for data science and numerical simulation can be of great help for these projects.
The Python open-source ecosystem could be equally leveraged as it is possible to bridge the two languages seamlessly using libraries such as PythonCall and JuliaCall \cite{pycall_juliacall}.
Most generally, \beforeit{} can contribute to the foundation of the next-generation tools for macro ABM research.

We foresee several lines of development that could further enhance \beforeit{}. 
One important enhancement is the modularisation and standardisation of the calibration scripts.
Unlike the rest of the package, these are not modularly written and currently come in the form of an unpolished research prototype. 
However, they could be easily improved and, potentially, even be adapted to make them work for any European nation. 
The resulting code could be integrated within \beforeit{} or be released as a dedicated companion package.
A second line of development concerns the further inclusion of models built on top of the \modelname{}. 
The package already ships two such variants, the \code{ModelCANVAS} variant inspired by the Bank of Canada's CANVAS model and the growth-rate \code{ModelGR} variant described here, and we plan to grow this library of variants in collaboration with the community.
Similarly, the multi-bank extension of the financial sector currently features banks that follow identical behavioural rules and differ only in their customer base.
Allowing banks to differ more fundamentally, for instance in their lending rates or capital requirements, would make \beforeit{} a natural tool for the analysis of systemic risk and macroprudential policy.
Finally, \beforeit{} is already compatible with automatic differentiation through the Mooncake compiler, which opens the way to gradient-based sensitivity analysis and calibration of the model to aggregate data. 
A natural next step is to broaden this support to additional AD backends, such as Enzyme, which would allow gradients to be cross-validated across independent implementations.
%

For any of the mentioned extensions or any other improvement to the package, we greatly encourage open-source contributions to the project in the form of issues, pull requests or feedback.


\section*{Acknowledgments}
We acknowledge open-source contributions from Peter Reschenhofer (Austrian Institute of Economic Research), Arnau Quera-Bofarull (Macrocosm) and Bence Mérő (Central Bank of Hungary) and we thank Claudia Biancotti (Banca d'Italia), Pietro Terna (University of Turin) and Marco Pangallo (University of Turin) for early feedback on the manuscript.
AG would like to thank the following colleagues from Banca d'Italia: Marco Benedetti, Claudia Biancotti and Marco Favorito for their continuous help, support and feedback on the project, Andrea Gentili and Sara Corbo respectively for their help on the design of the package's name and logo.
SP acknowledges funding from the Vienna Science and Technology Fund (WWTF) (grant number ESS22-040).
The views and opinions expressed in this paper are those of the authors and do not necessarily reflect the official policy or position of Banca d’Italia or those of the Bank of England or its committees.

\section*{Individual contributions}

AM greatly improved and extended the package including, among the many contributions, performance optimisations, testing infrastructure and support for automatic differentiation; 
MD improved the software quality of the first version of the package and designed the original extension macros; 
SH implemented the CANVAS model variant, growth-rate model variant, the forecasting significance tables, cross- and autocorrelation analysis, and the multicountry forecasting pipeline; 
OR developed the calibration to EU nations and the CalibrateBeforeIT.jl package; 
EM contributed to the code for multiple banks, testing and results output for multiple banks and conditional forecasting;
SP provided the original Matlab code and calibration scripts, and helped finding bugs in the original package; 
MB designed and implemented the infrastructure for multiple banks and the enhancements to CANVAS; 
AG conceived and coordinated the project, wrote the first version of the package, and wrote the initial manuscript, later improved with contributions from all authors.

\bibliographystyle{elsarticle-num}
\bibliography{refs}

@article{poledna2023economic,
  title={Economic forecasting with an agent-based model},
  author={Poledna, Sebastian and Miess, Michael Gregor and Hommes, Cars and Rabitsch, Katrin},
  journal={European Economic Review},
  volume={151},
  pages={104306},
  year={2023},
  publisher={Elsevier}
}

@article{hommes2025canvas,
  title={CANVAS: A Canadian behavioral agent-based model for monetary policy},
  author={Hommes, Cars and He, Mario and Poledna, Sebastian and Siqueira, Melissa and Zhang, Yang},
  journal={Journal of Economic Dynamics and Control},
  volume={172},
  pages={104986},
  year={2025},
  publisher={Elsevier}
}

@article{dawid2018agent,
  title={Agent-based macroeconomics},
  author={Dawid, Herbert and Delli Gatti, Domenico},
  journal={Handbook of computational economics},
  volume={4},
  pages={63--156},
  year={2018},
  publisher={Elsevier}
}

@article{axtell2025agent,
  title={Agent-based modeling in economics and finance: Past, present, and future},
  author={Axtell, Robert L and Farmer, J Doyne},
  journal={Journal of Economic Literature},
  volume={63},
  number={1},
  pages={197--287},
  year={2025},
  publisher={American Economic Association 2014 Broadway, Suite 305, Nashville, TN 37203-2425}
}

@article{dosi2019more,
  title={More is different... and complex! the case for agent-based macroeconomics},
  author={Dosi, Giovanni and Roventini, Andrea},
  journal={Journal of Evolutionary Economics},
  volume={29},
  pages={1--37},
  year={2019},
  publisher={Springer}
}

@article{kimmich2024economic,
  title={Economic impact of labor productivity losses induced by heat stress: an agent-based macroeconomic approach},
  author={Kimmich, Christian and Weyerstra{\ss}, Klaus and Czypionka, Thomas and Fauster, Norman FRM and Kinner, Maurice and Laa, Elisabeth and Mateeva, Liliana and Plank, Kerstin and Ulrici, Leonhard and Zenz, Hannes and Miess, Michael and Poledna, Sebastian},
  journal={Climatic Change},
  volume={178},
  number={3},
  pages={36},
  year={2025},
  publisher={Springer}
}

@article{hochrainer2024risk,
  title={Risk management against indirect risks from disasters: A multi-model and participatory governance framework applied to flood risk in {Austria}},
  author={Hochrainer-Stigler, Stefan and Bachner, Gabriel and Knittel, Nina and Poledna, Sebastian and Reiter, Karina and Bosello, Francesco},
  journal={International Journal of Disaster Risk Reduction},
  volume={106},
  pages={104425},
  year={2024},
  publisher={Elsevier}
}

@article{poledna2024economic,
  title={Economic and labour market impacts of migration in Austria: an agent-based modelling approach},
  author={Poledna, Sebastian and Strelkovskii, Nikita and Conte, Alessandra and Goujon, Anne and Linnerooth-Bayer, Joanne and Catalano, Michele and Rovenskaya, Elena},
  journal={Comparative Migration Studies},
  volume={12},
  number={1},
  pages={18},
  year={2024},
  publisher={Springer}
}

@article{bachner2024revealing,
  title={Revealing indirect risks in complex socioeconomic systems: A highly detailed multi-model analysis of flood events in {Austria}},
  author={Bachner, Gabriel and Knittel, Nina and Poledna, Sebastian and Hochrainer-Stigler, Stefan and Reiter, Karina},
  journal={Risk Analysis},
  volume={44},
  number={1},
  pages={229--243},
  year={2024},
  publisher={Wiley Online Library}
}

@misc{grazzini2023understanding,
  title={Understanding post-pandemic inflation dynamics with a behavioral macroeconomic model of the {Canadian} economy},
  author={Grazzini, Jakob and Hommes, Cars H and Poledna, Sebastian and Zhang, Yang},
  howpublished={SSRN Working Paper No. 4381235},
  year={2023},
  doi={10.2139/ssrn.4381235}
}

@article{hommes2023analyzing,
  title={Forecasting economic crises: The {Great Recession}, the sovereign debt crisis, and {COVID-19} in the euro area},
  author={Hommes, Cars and Poledna, Sebastian},
  journal={Economic Modelling},
  volume={158},
  pages={107497},
  year={2026},
  publisher={Elsevier}
}

@techreport{poledna2020recovery,
  title={Recovery of the {Austrian} economy following the {COVID-19} crisis can take up to three years},
  author={Poledna, S and Rovenskaya, E and Crespo Cuaresma, J and Kaniovski, S and Miess, M},
  year={2020},
  institution={International Institute for Applied Systems Analysis (IIASA)},
  type={IIASA Policy Brief},
  number={26}
}

@article{gill2021high,
  title={High-performance computing implementations of agent-based economic models for realizing 1:1 scale simulations of large economies},
  author={Gill, Amit and Lalith, Madegedara and Poledna, Sebastian and Hori, Muneo and Fujita, Kohei and Ichimura, Tsuyoshi},
  journal={IEEE Transactions on Parallel and Distributed Systems},
  volume={32},
  number={8},
  pages={2101--2114},
  year={2021},
  publisher={IEEE}
}

@inproceedings{sells2020julia,
  title={Julia programming language benchmark using a flight simulation},
  author={Sells, Ray},
  booktitle={2020 IEEE Aerospace Conference},
  pages={1--8},
  year={2020},
  organization={IEEE}
}

@techreport{ESA2010,
  title={European system of accounts - {ESA} 2010},
  author={Eurostat},
  type={Manuals and guidelines},
  institution={{Eurostat}},
  number={822},
  year={2013},
  howpublished={\url{https://ec.europa.eu/eurostat/web/products-manuals-and-guidelines/-/ks-02-13-269}}
}

@techreport{coletti2023blueprint,
  title={{A Blueprint for the Fourth Generation of Bank of Canada Projection and Policy Analysis Models}},
  author={Coletti, Donald},
  year={2023},
  institution={Bank of Canada},
  type={Staff Discussion Paper},
  number={{2023-23}}
}

@techreport{gosselin2023making,
  title={Making It Real: Bringing Research Models into Central Bank Projections},
  author={Gosselin, Marc-Andr{\'e} and Kozicki, Sharon},
  year={2023},
  institution={Bank of Canada},
  type={Staff Discussion Paper},
  number={{2023-29}}
}

@article{benedetti2022black,
  title={Black-it: A ready-to-use and easy-to-extend calibration kit for agent-based models},
  author={Benedetti, Marco and Catapano, Gennaro and De Sclavis, Francesco and Favorito, Marco and Glielmo, Aldo and Magnanimi, Davide and Muci, Antonio},
  journal={Journal of Open Source Software},
  volume={7},
  number={79},
  pages={4622},
  year={2022}
}

@inproceedings{glielmo2023reinforcement,
  title={Reinforcement Learning for Combining Search Methods in the Calibration of Economic {ABM}s},
  author={Glielmo, Aldo and Favorito, Marco and Chanda, Debmallya and Delli Gatti, Domenico},
  booktitle={Proceedings of the Fourth ACM International Conference on AI in Finance},
  pages={305--313},
  year={2023}
}

@inproceedings{brusatin2024simulating,
  title={Simulating the economic impact of rationality through reinforcement learning and agent-based modelling},
  author={Brusatin, Simone and Padoan, Tommaso and Coletta, Andrea and Delli Gatti, Domenico and Glielmo, Aldo},
  booktitle={Proceedings of the 5th ACM International Conference on AI in Finance},
  pages={159--167},
  year={2024}
}

@misc{Dynarejl,
  author       = {{Dynare Team}},
  title        = {Dynare.jl: A {Julia} rewrite of {Dynare} - solving, simulating and estimating {DSGE} models},
  year         = {2024},
  howpublished = {\url{https://github.com/DynareJulia/Dynare.jl}},
  note         = {Accessed: 2024-11-24}
}

@misc{StateSpaceEconjl,
  author       = {{Bank of Canada}},
  title        = {{StateSpaceEcon}.jl: A {Julia} package for working with Macroeconomic models},
  year         = {2024},
  howpublished = {\url{https://github.com/bankofcanada/StateSpaceEcon.jl}},
  note         = {Accessed: 2024-11-24}
}

@article{kockerols2023macromodelling,
  title={MacroModelling.jl: A {Julia} package for developing and solving dynamic stochastic general equilibrium models},
  author={Kockerols, Thore},
  journal={Journal of Open Source Software},
  volume={8},
  number={89},
  pages={5598},
  year={2023}
}

@techreport{del2015frbny,
  title={The {FRBNY DSGE} Model Meets {Julia}},
  author={Del Negro, Marco and Giannoni, Marc and Li, Pearl and Moszkowski, Erica and Smith, Micah},
  year={2015},
  institution={Federal Reserve Bank of New York}
}

@techreport{del2017forecasting,
  title={Forecasting with {Julia}},
  author={Del Negro, Marco and Giannoni, Marc and Gupta, Abhi and Li, Pearl and Moszkowski, Erica},
  year={2017},
  institution={Federal Reserve Bank of New York}
}

@misc{araujo2024open,
  title={Open-sourced central bank macroeconomic models},
  author={Araujo, Douglas},
  howpublished={SSRN Working Paper No. 4755247},
  year={2024},
  doi={10.2139/ssrn.4755247}
}

@techreport{turrell2016agent,
  title={Agent-based models: understanding the economy from the bottom up},
  author={Turrell, Arthur},
  institution={Bank of England},
  type={Quarterly Bulletin},
  number={Q4},
  year={2016}
}

@techreport{catapano2021macroprudential,
  title={Macroprudential policy analysis via an agent based model of the real estate sector},
  author={Catapano, Gennaro and Franceschi, Francesco and Loberto, Michele and Michelangeli, Valentina},
  type={Temi di Discussione (Working Paper)},
  institution={{Bank of Italy}},
  number={1338},
  year={2021}
}

@misc{iiasa-model-github,
  author       = {Sebastian Poledna and others},
  title        = {Economic forecasting with an agent-based model: Matlab code},
  year         = {2023},
  howpublished = {\url{https://github.com/iiasa/abm}},
  note         = {Accessed: 2024-11-27}
}

@misc{simudyne,
  author       = {Simudyne},
  title        = {Economic Forecasting Model},
  year         = {2024},
  howpublished = {\url{https://docs.simudyne.com/commercial_models/econ_forecast}},
  note         = {Accessed: 2025-01-13}
}

@inproceedings{hunold2020benchmarking,
  title={Benchmarking {Julia}’s communication performance: Is {Julia HPC} ready or Full {HPC}?},
  author={Hunold, Sascha and Steiner, Sebastian},
  booktitle={2020 IEEE/ACM Performance Modeling, Benchmarking and Simulation of High Performance Computer Systems (PMBS)},
  pages={20--25},
  year={2020},
  organization={IEEE}
}

@inproceedings{lin2021comparing,
  title={Comparing {Julia} to performance portable parallel programming models for {HPC}},
  author={Lin, Wei-Chen and McIntosh-Smith, Simon},
  booktitle={2021 International Workshop on Performance Modeling, Benchmarking and Simulation of High Performance Computer Systems (PMBS)},
  pages={94--105},
  year={2021},
  organization={IEEE}
}

@article{slepoy2008,
    author = {Slepoy, Alexander and Thompson, Aidan P. and Plimpton, Steven J.},
    title = {A constant-time kinetic {Monte Carlo} algorithm for simulation of large biochemical reaction networks},
    journal = {The Journal of Chemical Physics},
    volume = {128},
    number = {20},
    pages = {205101},
    year = {2008},
    month = {05},
    issn = {0021-9606}
}

@incollection{dawid2024implications,
  title={Implications of behavioral rules in agent-based macroeconomics},
  author={Dawid, Herbert and Delli Gatti, Domenico and Fierro, Luca Eduardo and Poledna, Sebastian},
  booktitle={The Economy as an Evolving Complex System {IV}},
  editor={del Rio-Chanona, R. M. and Pangallo, M. and Bednar, J. and Beinhocker, E. D. and Kaszowska-Mojsa, J. and Lafond, F. and Mealy, P. and Pichler, A. and Farmer, J. D.},
  pages={412--448},
  year={2026},
  publisher={SFI Press},
  address={Santa Fe, NM}
}

@article{d2022dynamic,
  title={Dynamic sampling from a discrete probability distribution with a known distribution of rates},
  author={D’Ambrosio, Federico and Bodlaender, Hans L and Barkema, Gerard T},
  journal={Computational Statistics},
  volume={37},
  number={3},
  pages={1203--1228},
  year={2022},
  publisher={Springer}
}

@misc{wiese2024forecasting,
  title={Forecasting macroeconomic dynamics using a calibrated data-driven agent-based model},
  author={Wiese, Samuel and Kaszowska-Mojsa, Jagoda and Dyer, Joel and Moran, Jose and Pangallo, Marco and Lafond, Francois and Muellbauer, John and Calinescu, Anisoara and Farmer, J Doyne},
  howpublished={arXiv preprint arXiv:2409.18760},
  year={2024},
  doi={10.48550/arXiv.2409.18760}
}

@misc{pycall_juliacall,
  author       = {{JuliaPy}},
  title        = {{PythonCall \& JuliaCall}},
  year         = {2024},
  howpublished = {\url{https://github.com/JuliaPy/PythonCall.jl}},
  note         = {Accessed: 2024-12-16}
}

@incollection{del2013dsge,
  title={{DSGE} model-based forecasting},
  author={Del Negro, Marco and Schorfheide, Frank},
  booktitle={Handbook of economic forecasting},
  volume={2},
  pages={57--140},
  year={2013},
  publisher={Elsevier}
}

@incollection{christiano2010dsge,
  title={{DSGE} models for monetary policy analysis},
  author={Christiano, Lawrence J and Trabandt, Mathias and Walentin, Karl},
  booktitle={Handbook of monetary economics},
  volume={3},
  pages={285--367},
  year={2010},
  publisher={Elsevier}
}

@book{hamill2015agent,
  title={Agent-based modelling in economics},
  author={Hamill, Lynne and Gilbert, Nigel},
  year={2015},
  publisher={John Wiley \& Sons}
}

@article{christiano2018dsge,
  title={On {DSGE} models},
  author={Christiano, Lawrence J and Eichenbaum, Martin S and Trabandt, Mathias},
  journal={Journal of Economic Perspectives},
  volume={32},
  number={3},
  pages={113--140},
  year={2018},
  publisher={American Economic Association 2014 Broadway, Suite 305, Nashville, TN 37203-2418}
}

@incollection{pangallo2024data,
  title={Data-driven economic agent-based models},
  author={Pangallo, Marco and del Rio-Chanona, R Maria},
  booktitle={The Economy as an Evolving Complex System {IV}},
  editor={del Rio-Chanona, R. M. and Pangallo, M. and Bednar, J. and Beinhocker, E. D. and Kaszowska-Mojsa, J. and Lafond, F. and Mealy, P. and Pichler, A. and Farmer, J. D.},
  pages={308--341},
  year={2026},
  publisher={SFI Press},
  address={Santa Fe, NM}
}

@misc{hafner2025sampling,
  author = {Hafner, Lilith Orion and Meligrana, Adriano},
  title = {Exact and efficient sampling from dynamic discrete distributions with finite-precision weights},
  howpublished = {arXiv preprint arXiv:2506.14062},
  year = {2025},
  doi = {10.48550/arXiv.2506.14062}
}

@article{datseris2024agents,
  title={Agents.jl: a performant and feature-full agent-based modeling software of minimal code complexity},
  author={Datseris, George and Vahdati, Ali R. and DuBois, Timothy C.},
  journal={Simulation},
  volume={100},
  number={10},
  pages={1019--1031},
  year={2024},
  publisher={Sage Publications Sage UK: London, England}
}

@inproceedings{moses2020autodiff,
 author = {Moses, William and Churavy, Valentin},
 booktitle = {Advances in Neural Information Processing Systems},
 editor = {H. Larochelle and M. Ranzato and R. Hadsell and M. F. Balcan and H. Lin},
 pages = {12472--12485},
 publisher = {Curran Associates, Inc.},
 title = {Instead of Rewriting Foreign Code for Machine Learning, Automatically Synthesize Fast Gradients},
 volume = {33},
 year = {2020}
}

@inproceedings{moses2021autodiff,
 author = {Moses, William S. and Churavy, Valentin and Paehler, Ludger and H\"{u}ckelheim, Jan and Narayanan, Sri Hari Krishna and Schanen, Michel and Doerfert, Johannes},
 title = {Reverse-Mode Automatic Differentiation and Optimization of GPU Kernels via Enzyme},
 year = {2021},
 isbn = {9781450384421},
 publisher = {Association for Computing Machinery},
 address = {New York, NY, USA},
 booktitle = {Proceedings of the International Conference for High Performance Computing, Networking, Storage and Analysis},
 articleno = {61},
 numpages = {16},
 pages={1--16},
 location = {St. Louis, Missouri},
 series = {SC '21}
}

@misc{innes2019autodiff,
  title={A differentiable programming system to bridge machine learning and scientific computing},
  author={Innes, Mike and Edelman, Alan and Fischer, Keno and Rackauckas, Chris and Saba, Elliot and Shah, Viral B and Tebbutt, Will},
  howpublished={arXiv preprint arXiv:1907.07587},
  year={2019},
  doi={10.48550/arXiv.1907.07587}
}

@misc{tebbutt2024autodiff,
  author = {Tebbutt, Will and Ge, Hong},
  title = {Mooncake: Towards a Differentiable General-Purpose Language},
  year = {2024},
  howpublished = {\url{https://github.com/chalk-lab/Mooncake.jl}},
  note = {Accessed: 2026-10-07}
}

@misc{revels2016autodiff,
  title={Forward-mode automatic differentiation in {Julia}},
  author={Revels, Jarrett and Lubin, Miles and Papamarkou, Theodore},
  howpublished={arXiv preprint arXiv:1607.07892},
  year={2016},
  doi={10.48550/arXiv.1607.07892}
}

@article{arya2022autodiff,
  title={Automatic differentiation of programs with discrete randomness},
  author={Arya, Gaurav and Schauer, Moritz and Sch{\"a}fer, Frank and Rackauckas, Christopher},
  journal={Advances in neural information processing systems},
  volume={35},
  pages={10435--10447},
  year={2022}
}

@inproceedings{andelfinger2021differentiable,
  title={Differentiable agent-based simulation for gradient-guided simulation-based optimization},
  author={Andelfinger, Philipp},
  booktitle={Proceedings of the 2021 ACM SIGSIM Conference on Principles of Advanced Discrete Simulation},
  pages={27--38},
  year={2021}
}

@misc{querabofarull2025automatic,
  title={Automatic differentiation of agent-based models},
  author={Quera-Bofarull, Arnau and Bishop, Nicholas and Dyer, Joel and Jarne Ornia, Daniel and Calinescu, Anisoara and Farmer, J Doyne and Wooldridge, Michael},
  howpublished={arXiv preprint arXiv:2509.03303},
  year={2025},
  doi={10.48550/arXiv.2509.03303}
}

@book{dennis1996numerical,
  title={Numerical Methods for Unconstrained Optimization and Nonlinear Equations},
  author={Dennis, John E. and Schnabel, Robert B.},
  series={Classics in Applied Mathematics},
  volume={16},
  publisher={SIAM},
  year={1996}
}

@book{nocedal2006numerical,
  title={Numerical Optimization},
  author={Nocedal, Jorge and Wright, Stephen J.},
  edition={2nd},
  publisher={Springer},
  year={2006}
}

@article{rumelhart1986learning,
  title={Learning representations by back-propagating errors},
  author={Rumelhart, David E and Hinton, Geoffrey E. and Williams, Ronald J.},
  journal={Nature},
  volume={323},
  number={6088},
  pages={533--536},
  year={1986},
  publisher={Nature Publishing Group UK London}
}

@article{baydin2018automatic,
  title={Automatic differentiation in machine learning: a survey},
  author={Baydin, Atilim Gunes and Pearlmutter, Barak A and Radul, Alexey Andreyevich and Siskind, Jeffrey Mark},
  journal={Journal of machine learning research},
  volume={18},
  number={153},
  pages={1--43},
  year={2018}
}

@inproceedings{dyer2023gradient,
  title={Gradient-assisted calibration for financial agent-based models},
  author={Dyer, Joel and Quera-Bofarull, Arnau and Chopra, Ayush and Farmer, J. Doyne and Calinescu, Anisoara and Wooldridge, Michael},
  booktitle={Proceedings of the Fourth ACM International Conference on AI in Finance},
  pages={288--296},
  year={2023}
}

@article{kotthoff2022calibrating,
  title={Calibrating agent-based models of innovation diffusion with gradients},
  author={Kotthoff, Florian and Hamacher, Thomas},
  journal={Journal of Artificial Societies and Social Simulation},
  volume={25},
  number={3},
  year={2022}
}

@article{kaplan2018monetary,
  title={Monetary policy according to HANK},
  author={Kaplan, Greg and Moll, Benjamin and Violante, Giovanni L.},
  journal={American Economic Review},
  volume={108},
  number={3},
  pages={697--743},
  year={2018},
  publisher={American Economic Association 2014 Broadway, Suite 305, Nashville, TN 37203}
}

@article{gali2007understanding,
  title={Understanding the effects of government spending on consumption},
  author={Gal{\'\i}, Jordi and L{\'o}pez-Salido, J David and Vall{\'e}s, Javier},
  journal={Journal of the european economic association},
  volume={5},
  number={1},
  pages={227--270},
  year={2007},
  publisher={Oxford University Press}
}

@article{carro2023heterogeneous,
  title={Heterogeneous effects and spillovers of macroprudential policy in an agent-based model of the {UK} housing market},
  author={Carro, Adrian and Hinterschweiger, Marc and Uluc, Arzu and Farmer, J. Doyne},
  journal={Industrial and Corporate Change},
  volume={32},
  number={2},
  pages={386--432},
  year={2023},
  publisher={Oxford University Press}
}

@article{bardoscia2025impact,
  title={The impact of prudential regulation on the {UK} housing market and economy: Insights from an agent-based model},
  author={Bardoscia, Marco and Carro, Adrian and Hinterschweiger, Marc and Napoletano, Mauro and Popoyan, Lilit and Roventini, Andrea and Uluc, Arzu},
  journal={Journal of Economic Behavior \& Organization},
  volume={229},
  pages={106839},
  year={2025},
  publisher={Elsevier}
}

@misc{calibratebeforeitjl,
  author       = {{Vienna Institute for International Economic Studies (wiiw)}},
  title        = {{CalibrateBeforeIT.jl}: Calibration functions for the {BeforeIT.jl} {ABM}},
  year         = {2026},
  howpublished = {\url{https://github.com/ViennaInstitute/CalibrateBeforeIT.jl}},
  note         = {Accessed: 2026-10-07}
}

@techreport{minnella2026devil,
  title={The devil in the {DeTail}: assessing state-contingent tail effects of a releasable macroprudential capital buffer using a parsimonious agent-based framework},
  author={Minnella, Enrico and Pereira, Ana and Tereanu, Eugen},
  institution={{Bank of England}},
  type={Staff Working Paper},
  number={1198},
  year={2026}
}

\end{document}